\documentclass{article} 

\usepackage[utf8]{inputenc} 

\usepackage{geometry} 
\usepackage{graphicx} 

\usepackage{booktabs} 
\usepackage{array} 
\usepackage{paralist} 
\usepackage{verbatim} 
\usepackage{subfig} 

\usepackage{fancyhdr} 
\usepackage{sectsty}
\allsectionsfont{\sffamily\mdseries\upshape} 

\usepackage[nottoc,notlof,notlot]{tocbibind} 
\usepackage[titles,subfigure]{tocloft} 

\title{Conservation Laws in a Limit Order Book}
\author{Jan Rosenzweig\\ Pine Tree Funds, j.rosenzweig@pinetree-funds.com\footnote{JR is a Fund Manager at Pine Tree and a Visiting Senior Research Fellow at King's College London, jan.rosenzweig@kcl.ac.uk}}
\date{} 

\begin{document}
\maketitle

\abstract{We present a class of macroscopic models of the Limit Order Book to simulate the aggregate behaviour of market makers in response to trading flows. The resulting models are solved numerically and asymptotically, and a class of similarity solutions linked to order book formation and recovery is explored. The main result is that order book recovery from aggressive liquidity taking follows a simple $t^{1/3}$  scaling law.}

\section{Introduction}

Limit order book is at the heart of trading on any organised venue. Dynamics of the order book is central to understanding the price and liquidity impact of any trading strategy, and it is important in a number of intraday trading strategies. 

Several models of the orderbook dynamics have been proposed over the years, most notably \cite{AlfonsiFruthSchied}, \cite{AlfonsiSchiedSlynko}, \cite{Boucheaud}, \cite{ObizhaevaWang}, \cite{Yura} and others. These models generally assume a generic shape of the order book, and a process for its recovery when liquidity is taken out through trading. 

We take a slightly different approach, inspired by concepts from fluid dynamics. Separating the order book dynamics into conserved and non-conserved components, we get a generic framework into which we can plug in increasingly complex assumptions about interactions between limit orders in the book. 

The result is a model which is sufficiently simple to allow analytical insight, and sufficiently realistic to provide a number of features of real life order books.

We are specifically interested in processes governing order book formation and recovery, and we look at some simple solutions to the resulting transport equation which are relevant in that context.

\section{Basic dynamics of the Limit Order Book}

\begin{figure}[h]
  \includegraphics[scale=0.6]{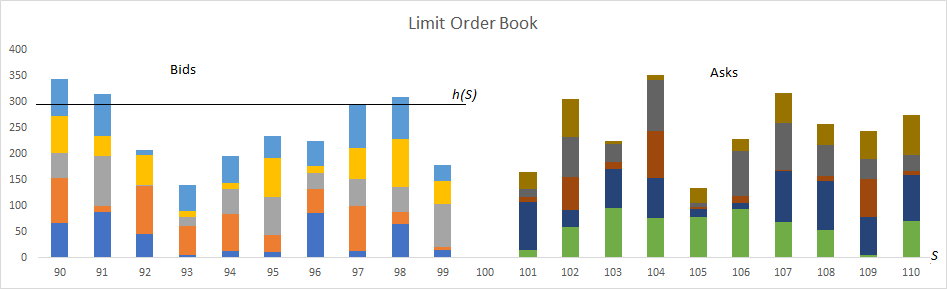}
  \caption{Limit Order Book}
  \label{fig:lob}
\end{figure}

Simple schematics of the limit order book is shown in Figure \ref{fig:lob}. We denote by $S$ the price level, and by $h(S,t )$ the total quantity of limit orders available at level $S$ at time $t$. We use subscript to denote partial differentiation, so $h_{t}= \partial h/\partial t$ and $h_{S}=\partial h/\partial S$. 

\subsection{Conservation laws}

The quantity between two price levels $S_{1}$ and $S_{2}$ is given by 

\begin{equation}
V(S_{1},S_{2},t) = \int_{S_{1}}^{S_{2}} h(S,t)\ dS \label{volume}
\end{equation}
and the quantity flux is defined as 
\begin{equation}
Q(S_{1},S_{2},t) =  \frac{d}{dt} V(S_{1},S_{2},t), \label{flux}
\end{equation}
the instantaneous rate of flow of quantity through the given part of the order book.
Then, differentiating (\ref{volume}) with respect to time and using  integration by parts, we get the \textit{law of conservation of quantity} in the form 
\begin{equation}
h_{t}(S,t) - Q_{S} (S,t) = P(S,t) \label{cons}
\end{equation}
where $Q(S,t)$ is taken to mean $Q(-\infty,S,t)$ for the bid side, and $Q(S,\infty,t)$ for the ask side. 

Note that it is unusual to make the underlying dynamics lognormal in orderbook models, since the effects of lognormality are tiny on relevant price and time scales. Hence we formally allow bids lower than zero, but  this has no meaningful effects on the analysis.

Equation (\ref{cons}) is analogous to the standard physical conservation of mass. The left hand side contains the conserved terms, or flow of quantity through the orderbook, and the right hand side denotes any creation or destruction of quantity in the orderbook. Equation (\ref{cons}) has a simple  interpretation, namely that any changes of quantity at a given price level (term $h_{t}$) correspond to quantity migrating between levels (term $Q_{S}$), or quantity being added to or taken out of the order book  (term $P$).

The source/sink term $P$ does not correspond to quantity being extinguished through aggressive liquidity taking, which would be handled through  boundary condition to the equation (\ref{cons}), but it purely corresponds to quantity changes away from the touch.

The case where orders can be created or cancelled, but quantity is never moved between levels encompasses a number of well known order book models. If there is no quantity flow between levels, then $Q \equiv 0$ and 
equation (\ref{cons}) becomes
\begin{equation}
h_{t}(S,t)  = P(S,t) \label{classic}
\end{equation}
This formulation includes all of the models from \cite{AlfonsiFruthSchied}, \cite{AlfonsiSchiedSlynko}, \cite{Gatheral}, \cite{GatheralSchiedSlynko}, \cite{ObizhaevaWang}, \cite{Toth}, \cite{Weiss}, \cite{Yura}, and many others.

\subsection{Flowing orders}

To analyse the flow of orders inside the order book, we borrow a few ideas from fluid dynamics. We introduce the cocnept of \textit{velocity} $u(S,q,t)$ of volume at a price level $S$, orderbook priority queue $q$ and time $t$. The velocity $u$ is related to the flow rate $Q$ through
\begin{equation}
Q(S,t)  = \int_{0}^{h(S,t)} u(S,q,t)\ dq \label{fluxvel}
\end{equation}
and it has the interpretation of the rate at which quantity flows from one level to another.

Having introduced the model in which quantity is free to flow between levels, we now need a constitutive equation to provide a microstructural mechanism driving such movement.
Continuing the fluid dynamical analogy, we introduce the concept of \textit{pressure} $p(S,q,t)$ at a price level $S$ at time $t$ driving the flow of quantity.

We can now formulate the constitutive equation. We assume that pressure is uniform within each level ($p(S,q,t)=p(S,t)$ and that it is proportional to the occupancy of the level,
\begin{equation}
p(S,t)  = \theta h(S,t) \label{gravity}
\end{equation}
for some constant coefficient $\theta$.

The interpretation of equation (\ref{gravity}) is that traders look to avoid adverse selection by looking for empty or near-empty levels. The more quantity is already available at the level, the higher the pressure at that level, and the lower the likelihood of new quantity being added to that level.

\begin{figure}[h]
  \includegraphics{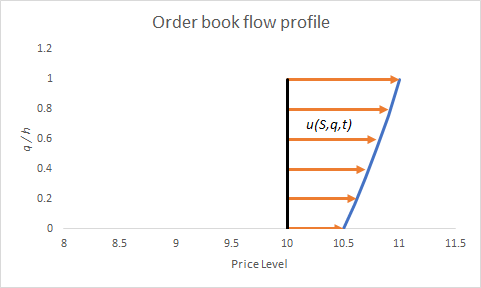}
  \caption{Quantity flow profile in the order book}
  \label{fig:profile}
\end{figure}

To that, we still have to add another constitutive equation providing the velocity profile. We use an algebraic velocity profile of the form
\begin{equation}
u(S,q,t)  = \frac{1}{\rho} p_{S}(S,t) \left[ \left( \frac{q}{h(S,t)} \right)^{\beta} + u_{0}(S,t) \right]  \label{euler}
\end{equation}
for some constants $\rho, \beta > 0$, as shown in Figure \ref{fig:profile}. The interpretation of (\ref{euler}) is that the force trying to move orders  is proportional to the pressure gradient (term $p_{S}$), while the ability of orders to be moved depends on their relative place in the queue at their level. Orders at the front of the queue ($q=0$) are relatively immobile, and they can only move at some slip rate $u_{0}$. The orders at the end of the queue are most mobile, and they are most likely to move around to avoid adverse selection. Finally, the division by $h$ ensures that the dependence on the place in the queue is relative rather than absolute. In other words, an order with $q$ lots in front of it may be either very immobile if there is a long queue behind it ($q \ll h$), or very mobile if it is the last lot in the queue ($q=h$).

Putting equations (\ref{cons}), (\ref{fluxvel}), (\ref{gravity}) and (\ref{euler}) together, we get
\begin{equation}
h_{t}(S,t)  =  \frac{\rho}{ \theta(\beta + 1)}  \left[ h(S,t)\ h_{S}(S,t) + u_{0} h(S,t) \right]_{S} + P(S,t). \label{inviscid1}
\end{equation}
Finally, we can group the constants into a single time scale to which we rescale the time to get the transport equation
\begin{equation}
h_{t}(S,t) =     h^{2}(S,t)_{SS} + \left( u_{0}h(S,t) \right)_{S}  + P(S,t),  \label{inviscid}
\end{equation}
where $t$ now denotes the rescaled time, $t\mapsto t/ \theta(\beta + 1)$.

The boundary conditions usually take one of the forms
\begin{equation}
\begin{array}{ccc}
h(S_{0},t)=h_{0}; & h_{S}(S_{0},t) = p_{0}; & h^{2}(S_{0},t)_{S} = Q_{0}
\end{array}
\end{equation}
The boundary conditions specify the orderbook thickness, slope or flux at the level $S_{0}$, respectivel, and $S_{0}$ denotes an appropriate price level. $S_{0}$ is usually, but not necessarily, chosen as either the touch level, or some appropriate price level inside the spread. 

In most practical cases, the slip term can be set to zero, $u_{0}=0$.

The transport equation  (\ref{inviscid})  is well known in fluid dynamics, where it describes the propagation of gravity driven thin layer of an inviscid fluid. The derivation in the fluid dynamics case is somewhat different, in that inviscid fluid has a steady profile which is not normalised to film height $h$, $u(S,q,t) = u(S,t)$.

Transport equation is a second order, nonlinear isotropic diffusion equation. Isotropy is sometimes considered to be a problem, because flow of orders towards the touch level is not the same as the flow of orders away from the touch levels. On the other hand, anisotropy can be provided through the boundary conditions. Intuitively, flows in deep book are generally dominated by competition for queue priority, and they are at leading order isotropic. 

\section{Solving the Transport Equation}

\subsection{Steady state and the $S^{1/2}$ profile}

Setting $u_{0} \equiv P \equiv 0$, we can find a family of steady state solutions of the form 
\begin{equation}
h(S,t) = \pm  \left| \pm a^{2} (S - S_{b} )\right|^{1/2} \label{steady}
\end{equation}
for some $a,S_{b}$.

\begin{figure}[h]
  \includegraphics{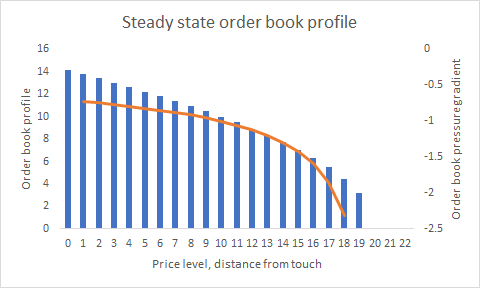}
  \caption{Square root steady state profile of the ask order book. Quantities are shown as blue bars, pressure is shown as amber line. Price levels are shown as distance from the touch.}
  \label{fig:steady}
\end{figure}

The only set of solutions with finite total order book quantity is the one with negative $\pm$ sign inside the square root, and it implies a decaying square root profile maximised at the touch level, and zero beyond the point $S_{b}$, as shown in figure \ref{fig:steady}.

In this solution, the pressure is maximised at the touch level, and pressure falls off in deep book. The book thickness reaches zero after a finite number of levels. 

While this is not a particularly interesting family of solutions in its own right, it serves as a set of  long time attractors for various solutions. One of the parameters in (\ref{steady}) can be eliminated through the requirement that the orderbook quantity is retained, but this still leaves a one-parameter family of solutions. This remaining parameter is usually determined from the boundary conditions.

\subsection{Similarity solutions and the $t^{1/3}$ scaling}

We look for  more interesting solutions in the similarity form,
\begin{equation} h(S,t) =  H(t) v(s),\ s = \frac{S-S_{0}}{L(t)} \label{similar}
\end{equation}

All terms are balanced if the scaling is  
\begin{equation}
H(t) \propto t^{-1/3},\ L(t) \propto t^{1/3},\  S_{0}(t) \propto t^{1/3} \label{scaling}
\end{equation}
and the similarity function $v$ then satisfies
\begin{equation}
3 ( v^{2}) '' +  v's - \gamma v' +  v = 0 \label{similarity}
\end{equation}
\begin{equation}
v(0) = v(\infty) = 0 \label{similaritybc}
\end{equation}
where 
\begin{equation}
\gamma = -\frac{H}{L\dot{H}} \dot{S_{0}}
\label{gamma}
\end{equation}
is the dimensionless speed of the touch level. The minus sign, combined with the substitution (\ref{similar}) keeps the intuitive distinction that $\gamma > 0$ corresponds to the advancing touch level (increasing best bid, or decreasing best ask), and $\gamma < 0$ corresponds to retreating touch level (decreasing best bid, or increasing best ask). If the touch level is neither advancing nor retreating, $\gamma=0$.

Equations (\ref{similarity}), (\ref{similaritybc}) do not  have analytical solutions, but they can be solved asymptotically and numerically.

Around the touch, the leading order asymptotics  for $\gamma \ge 0$ is
\begin{equation}
v(s) \approx \frac{\gamma}{6} s -\frac{1}{4} s^{2} + \frac{\gamma}{118}s^{3} + O(s^{4}). \label{touch}
\end{equation}

There is no positive similarity solution for $\gamma<0$.

In the deep book, we find
\begin{equation}
v(s) \approx v_{\infty} \left( \frac{1}{s} +  \frac{\gamma}{s^{2}} +  \frac{\gamma^{2}}{s^{3}}+  O(s^{-4}) \right) . \label{farfield}
\end{equation}
for some deep book constant $v_{\infty}>0$. The far field decays like $h \propto 1/s$ at leading order, with the motion of the touch only providing a higher order correction. Note that the similarity solution formally implies infinite order book volume, due to its deep book decay being only $1/s$. In practice, there are no fractional quantities in a real life order books, so the corresponding sum would be finite in real life.

The similarity equation (\ref{similarity}) is symmetric for $\gamma=0$, hence at leading orders symmetric terms are independent of $\gamma$ near the touch (\ref{touch}), while the antisymmetric terms depend strongly on $\gamma$ and vanish for $\gamma=0$. At higher orders, the effect of $\gamma$ begins to mix into symmetric terms due to nonlinearity of (\ref{similarity}), and the mixing is obvious in the deep book limit (\ref{farfield}). 

Substituting (\ref{touch}), (\ref{farfield}) back into (\ref{inviscid}), we get the touch similarity solution $h_{0}$ and the deep book similarity solution $h_{\infty}$ as 
\begin{equation}
h_{0}(S,t) \approx \frac{S-S_{0}}{ t^{2/3}} \left[ \frac{\gamma}{6}  -\frac{1}{4} \frac{S-S_{0}}{ t^{1/3}}  + \frac{\gamma}{118}\frac{(S-S_{0})^{2}}{ t^{2/3}} + O\left(\frac{(S-S_{0})^{4}}{ t}\right) \right], \label{fulltouch}
\end{equation}
\begin{equation}
h_{\infty}(S,t) \approx v_{\infty} \left( \frac{1}{S-S_{0}} +  \frac{\gamma t^{1/3}}{(S-S_{0})^{2}} +  \frac{\gamma^{2}t^{2/3}}{(S-S_{0})^{3}}+   O\left(\frac{ t}{(S-S_{0})^{4}}
\right)\right) . \label{fullfarfield}
\end{equation}

\begin{figure}[h]
  \includegraphics{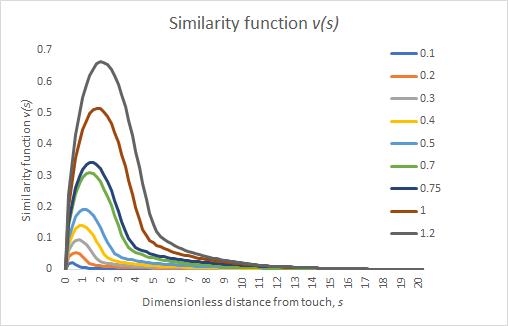}
  \includegraphics{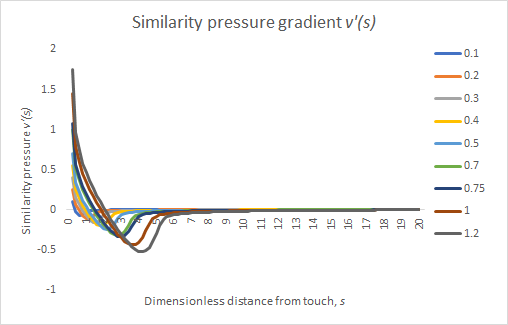}
  \caption{Similarity function $v(s)$ and similarity pressure gradient $v'(s)$ for different values of dimensionless touch speed $\gamma$.}
  \label{fig:similarity}
\end{figure}

There is an entire family of similarity solutions satisfying the deep book boundary condition (\ref{farfield}), parametrized by the deep book constant $v_{\infty}$. Only one of them satisfies the touch condition (\ref{touch}) for any $\gamma$, and it has to be calculated numerically. The profiles of the similarity function for various values of touch speed / touch angle $\gamma$ are shown in Figure \ref{fig:similarity}.

The numerical solution for the similarity function behaves as expected from the asymptotics. It leaves the dimensionless touch at zero with finite angle $\gamma$, it peaks at some point beyond the dimensionless touch, and then it decays as $1/s$ into the deep book. Its assosiacted pressure peaks at the touch, and it is balanced by negative pressure at a point beyond the peak of the profile. The pressure then relaxes to uniform pressure in deep book. 

The interpretation is that  orders several layers behind the touch find a possibility to improve their queue position by migrating towards the touch, leaving their original place in the queue and generating negative presure there. The accumulaton of these competing orders near the touch generates a high degree of competition for a place in the queue, manifested as high pressure, and this moves the touch forward. Orders sufficiently deep in the book do not participate in this queue competition, and their pressure is uniform.

It is of note that the profile of the solutions to (\ref{similarity}) looks very similar to the order density of Gatheral \cite{Gatheral}, Bouchaud et al \cite{Boucheaud} and many others. This is not entirely surprising; this profile is largely dictated by the boundary conditions (\ref{similaritybc}); the simplest non-negative function subject to boundary conditions (\ref{similaritybc}) rises from $v(0)=0$, reaches a finite positive maximum, and then decays back to $v(\infty) \sim 0$.

\subsection{Simulations}

We have performed simulations for the advancing order book in two scenarios. In the first scenario, the ask order book was filled with uniform quantity above a fixed price level, and left to evolve. In the second, everything was the same but the touch was blocked from advancing below a fixed level, chosen to simulate the effect of the bid book. 

The effect we simulate is the recovery of the order book after a large execution has wiped out a number of ask levels. 

\begin{figure}[h]
\includegraphics[scale=0.6]{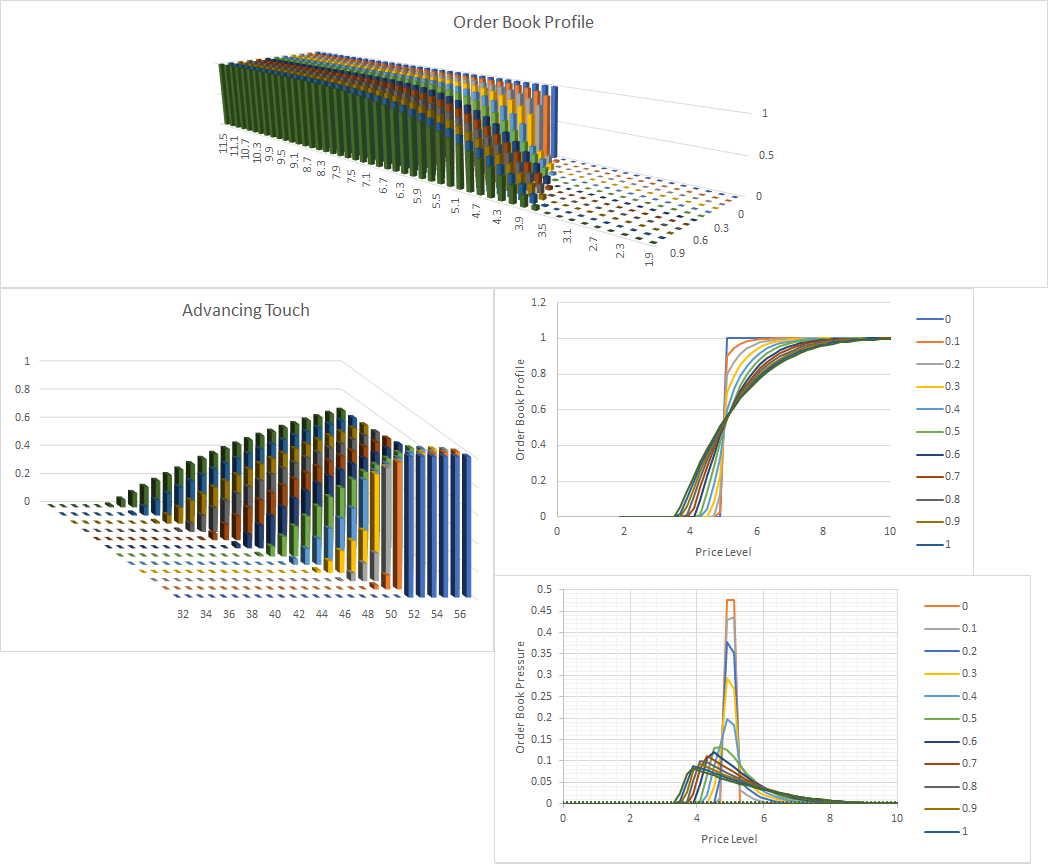}
\caption{Advancing Ask book, no limits. The touch pressure gradient is initially high due to the sharp cut-off, and it is quickly smoothed out. Subsequently, the self similar pressure wave propagates the touch level as $t^{1/3}$.}
\label{fig:unlimited}
\end{figure}

\begin{figure}[h]
\includegraphics[scale=0.6]{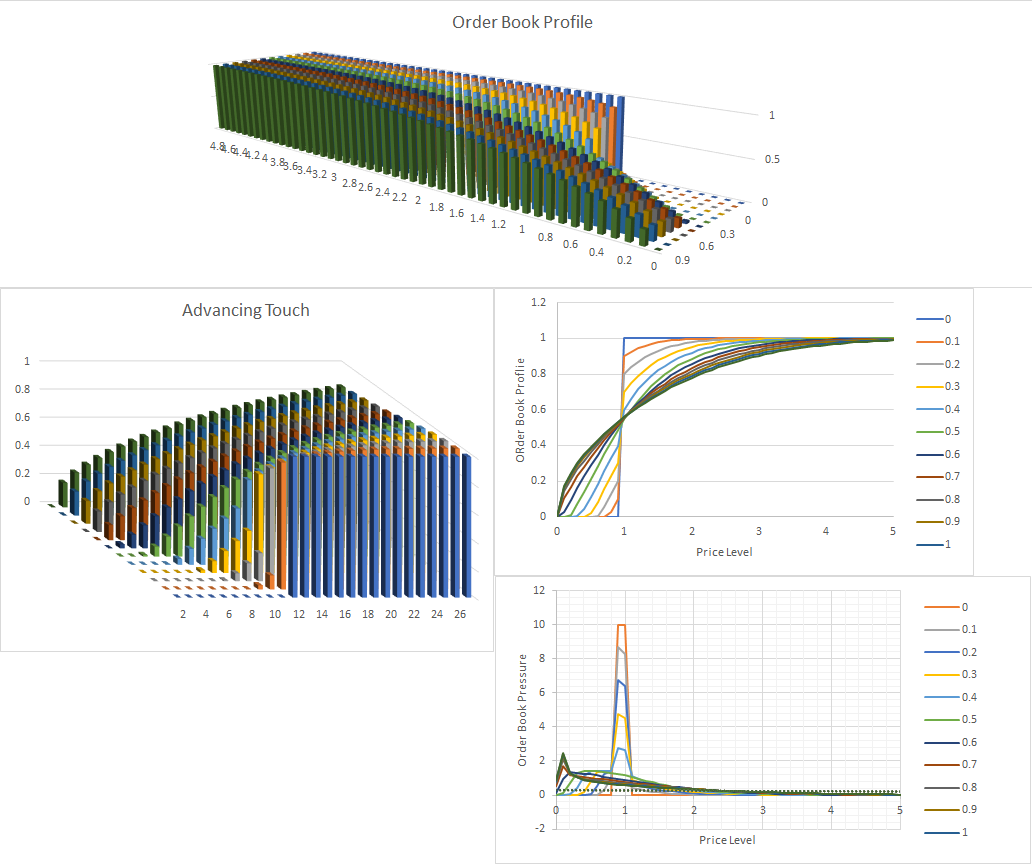}
\caption{Advancing Ask book, limited at fixed price level.The touch pressure gradient is initially high due to the sharp cut-off, and it is quickly smoothed out. Subsequently, the self similar pressure wave propagates the touch level as $t^{1/3}$. Once the limit has been reached, the pressure wave can propagate no further and pressure builds up at the constrained touch level. }
\label{fig:limited}
\end{figure}

The orderbook profiles and corresponding pressures for the two scenarios are shown in Figures \ref{fig:unlimited} and \ref{fig:limited}. 

The initial profile has a high pressure point due to the sharp cut-off at the initial touch level, and the initial stage consists of smoothing out this kink by pushing quantity both inside the spread and into deep book.

Once the effect of the initial cut off has worn out, the orderbook profile assumes the form of the self similar pressure wave advancing according to the $t^{1/3}$ similarity solution. 

In both examples, we have put a firm stop to advancing touch at the price level $S=0$, simulating the effect of the bid book. In the first example, the simulation was started sufficiently far from the firm stop so that its effects are not felt. In the second example, the simulation was started near the stop so that the stop is reached quickly.

The first example maintains the self similar pressure wave, and it would continue to maintain it for ever in the absence of a firm stop. It continues to spread ever smaller quantities deper and deeper into the spread according to the $t^{1/3}$ similarity solution.

The second example reaches the firm stop relatively soon after  the simulation was started. Self similar propagation ends once the firm stop has been hit, the touch level stops advancing, and pressure builds up at the touch level. In the limit, the pressure is maximised at the constrained touch level, and it falls off in deep book. The pressure gradient is imposed by the firm stop to the advancing contact line. The only way to releave the pressure at the touch would be to release the firm stop by crossing the spread from the ask side and taking out levels from the bid book, or by reducing the quantity at touch by crossing the spread from the bid side.

Once the similarity scaling breaks down due to the imposition of hard stop, the soluction begins slow convergence to the steady $S^{1/2}$ profile from Figure \ref{fig:steady}. This convergence is very slow, and it is of no practical importance for understanding real life order books.

\section{Conclusions}
We have presented a simple model of quantity flows in a limit order book. We are primarily interested in quantity-conserving flows, and we only use a simple assumption about the behaviour of market makers. Namely, we assume that market makers would rather join relatively empty levels than relatively full levels, and that orders' ability to move around depends on their place in the queue - the first order in the queue is the least mobile, and last order in the queue is the most mobile. We use an algebraic velocity profile, but the form of the resulting transport equation comes out independent of the choice of the algebraic profile. The choice of the algebraic profile only affects the resulting relaxation time scale.

The model is not intended to simulate a real world order book with all its complexities.  Instead, it is intended to simulate, in a statistical ensemble sense, the collective behaviour of market makers and their reactions to each other and other traders.

As such, it would be foolish to take a snapshot of any real world order book, feed it to our model as the initial condition, and expect a realistic prediction of the shape of the order book in the future. 

The model is, however, useful in isolating various effects that make up the behaviour of real life order books. The specific worked out example concerns the reaction of the order book after significant quantity has been taken out by aggressive traders, and the way the interaction between market makers decays the initial impact. 

We show that the relaxation is self similar with the $t^{1/3}$ scaling until the suitably chosen mid point ("mid point" being used in the most general sense) has been reached, building up a new steady state. The resulting steady state has high pressure at the thus formed touch level, and it is this high pressure that leads subsequent reactions of the order book to new situations.

There is a limiting steady state solution with $S^{1/2}$ profile , having peak pressure at the touch evel and square root decay to zero after finitely many levels. This steady state serves as a long time attractor to the solutions with fixed touch level. However, the time scales in which the steady state is reached are very long, and they exceed any reasonable time scales related to real life order books.

\end{document}